\documentclass[letterpaper, 10 pt, journal, twoside]{ieeetran}
\usepackage{amsmath, amssymb}
\usepackage{graphicx}
\usepackage{subfigure}
\usepackage[noadjust]{cite}
\usepackage[version=3]{mhchem}
\newtheorem{theorem}{Theorem}
\newtheorem{remark}{Remark}
\usepackage{color}
\usepackage{amsmath, amssymb}
\usepackage{graphicx}
\usepackage{subfigure}
\usepackage[noadjust]{cite}
\usepackage{color}
\usepackage[noadjust]{cite}
\usepackage{color}
\usepackage{wrapfig}
\usepackage{hyperref}
\usepackage{mathrsfs}
\usepackage{cite}
\IEEEoverridecommandlockouts                              

\title{Inverse Stability Problem and Applications to Renewables Integration}

\author{~Thanh Long Vu$^*$, Hung Dinh Nguyen, Alexandre Megretski, Jean-Jacques Slotine  and~Konstantin~Turitsyn
\thanks{$^*$Corresponding author. Email: longvu@mit.edu. All the authors are with the Massachusetts Institute of Technology, Cambridge, MA 02139, USA. 
}}

\begin{document}


\maketitle
\begin{abstract}
In modern power systems, the operating point, at which the demand and supply are balanced, \textcolor{black}{may take different values due to changes in loads} and renewable generation levels. Understanding the dynamics of stressed power systems with \textcolor{black}{a range of} operating points would be essential to \textcolor{black}{assuring} their reliable operation, and possibly allow higher integration of renewable \textcolor{black}{resources}. This letter introduces a non-traditional way to think about the stability assessment problem of power systems. Instead of estimating the set of initial states leading to a given operating condition, we characterize the set of operating conditions that a power grid converges to from a given initial state under changes in power injections and lines. We term this problem as ``inverse stability'', a problem which is rarely addressed in the control and systems literature, and hence, poorly understood. Exploiting quadratic \textcolor{black}{approximations} of the system's energy function, we introduce an estimate of the inverse stability region. Also, we briefly describe three important applications of the inverse stability notion: (i) robust stability assessment of power systems w.r.t. different renewable generation levels, (ii) stability-constrained optimal power flow (sOPF), and (iii) stability-guaranteed corrective action design. 
\end{abstract}

\maketitle
\pagenumbering{gobble}

\section{Introduction}

\IEEEPARstart{R}{enewable} generations, e.g., wind and solar, are increasingly installed into electric power grids to reduce \textcolor{black}{$\ce{CO_2}$} emission from the electricity generation sector. Yet, their natural
intermittency presents a major challenge to the delivery of
consistent power that is necessary for today's grid operation, in which generation must instantly meet load. 
Also, the inherently low inertia of renewable
generators limits the grid's controllability and makes \textcolor{black}{it easy for the grid to lose} its stability. The existing power grids and management tools were not designed to deal with these new challenges.
Therefore, new stability assessment and control design tools are needed to \textcolor{black}{adapt to the changes in architecture and dynamic behavior expected in the future power grids}.

Transient stability assessment of power system certifies that the system state converges to a stable operating condition after the system experiences large disturbances. Traditionally, this task is handled by using either time domain simulation (e.g., \cite{Nagel:2013kf}), or by utilizing the energy method (e.g., \cite{pai2012energy,  Chiang:2011eo}) and the Lyapunov function method  (e.g., \cite{Hiskens:1997Lya}) to estimate the stability region of a given equilibrium point (EP), i.e., the set of initial states from which the system state will converge to that EP.
In modern renewable power grids, the operating point \textcolor{black}{may take different values under} the real-time clearing of electricity markets, intermittent renewable generations, changing loads, and external disturbances. \textcolor{black}{Dealing with the situation when the EP can change over a wide range} makes the transient stability assessment even more technically difficult and computationally cumbersome.

In this letter, rather than considering the classical stability assessment problem, we formulate the \emph{inverse stability assessment problem}. This problem concerns with estimating the region around a given initial state \textcolor{black}{$\delta_0$}, called \emph{``inverse stability region''} \textcolor{black}{$\mathcal{A}(\delta_0)$}, so that whenever the power injections or power lines change and \textcolor{black}{lead to an} EP in \textcolor{black}{$\mathcal{A}(\delta_0)$}, the system state will converge from  \textcolor{black}{$\delta_0$} to that EP.  Indeed, the convergence from $\delta_0$ to an EP is guaranteed when the system's energy function is bounded under some threshold \cite{pai2012energy,Chiang:2011eo}. In \cite{VuTuritsyn:2015TAC}, we observed that if the EP \textcolor{black}{is in the interior of the set} $\mathcal{P}$ characterized by phasor angular differences smaller than $\pi/2,$ then the nonlinear power flows can be strictly bounded by linear functions of angular differences. Exploiting this observation, we show that the energy function of power system can be approximated by quadratic functions of the EP and the system state, and from which we obtain an estimate of the inverse stability region.

The remarkable advantage of the inverse stability certificate is \textcolor{black}{making it possible} to exploit the change in EP to achieve useful dynamical properties. We will briefly discuss three applications of this certificate, which are of importance to the integration of large-scale renewable resources:
\begin{itemize}
\item []{\bf Robust stability assessment:} For a typical power system composed of \textcolor{black}{several components and integrated with different levels of renewable generations}, there are many contingencies that need to be  reassessed on a regular basis. Most of these contingencies correspond to failures of relatively small and insignificant components, so the post-fault dynamics is probably transiently stable. Therefore, most of the computational effort is spent on the analysis of non-critical scenarios. This computational burden could be greatly alleviated by a robust transient stability assessment toolbox that could certify the system's stability w.r.t. a broad range of uncertainties. In this letter, we show that the inverse stability certificate can be employed to assess the transient stability of power systems for various levels of power injections. 

\item []{\bf Stability-constrained OPF:} Under large disturbances, a power system with \textcolor{black}{an operating condition derived by solving the conventional OPF problem} may not survive. It is therefore desirable to design operating conditions  \textcolor{black}{so that the system} can withstand large disturbances. This can be carried out by incorporating the transient stability constraint into OPF together with the normal voltage and thermal constraints. Though this problem was discussed in the literature \textcolor{black}{(e.g., \cite{Pizano-Martianez2010})}, there is no way to precisely formulate and solve the stability-constrained OPF problem because transient stability is a dynamic concept and differential equations
are involved in the stability constraint.
Fortunately, the inverse stability certificate allows for a natural incorporation of the stability constraint into the OPF problem as a static constraint  of placing the EP in a given set. 

\item []{\bf Stability-guaranteed corrective actions:} 
Traditional protection strategies focus on the safety of individual components, and the level of coordination among component protection systems is far from perfect. Also, they do not take full advantage of the new flexible and fast electronics resources available in modern power systems, and largely rely on customer-harmful actions like load shedding.  These considerations motivated us to coordinate widespread flexible electronics resources as a system-level \textcolor{black}{customer-friendly} corrective action with guaranteed stability \cite{vu2016structural}. This letter presents a unconventional control way in which we relocate the operating point, by appropriately redispatching power injections, to attract the emergency state and stabilize the power systems under emergency situations.  
\end{itemize}


\section{Inverse stability problem of power systems}
\label{sec.model}

In this letter, we utilize the  
structure-preserving model to describe the power system dynamics \cite{bergen1981structure}.  This model naturally
incorporates the dynamics of the generators' rotor angle and the response of
load power output to frequency deviation. 
Mathematically, the grid is described by an undirected graph
$\mathcal{A}(\mathcal{N},\mathcal{E}),$ where
$\mathcal{N}=\{1,2,\dots,|\mathcal{N}|\}$ is the set of buses and
$\mathcal{E} \subseteq \mathcal{N} \times \mathcal{N}$ is the set
of transmission lines \textcolor{black}{$\{k,j\}, k,j \in \mathcal{N}$}. Here, $|A|$ denotes
the number of \textcolor{black}{elements of set $A$}. The sets of generator buses
and load buses are denoted by $\mathcal{G}$ and $\mathcal{L}$. We assume that the grid is lossless with
constant voltage magnitudes $V_k, k\in \mathcal{N},$ and the
reactive powers are ignored. Then, the grid's dynamics is described by \cite{bergen1981structure}:
\begin{subequations}
\label{eq.structure-preserving}
\begin{align}
\label{eq.structure-preserving1}
 m_k \ddot{\delta_k} + d_k \dot{\delta_k} + \sum_{\{k,j\} \in
  \mathcal{E}} a_{kj} \sin(\delta_k-\delta_j) = &P_{k},  k \in \mathcal{G},  \\
  \label{eq.structure-preserving2}
  d_k \dot{\delta_k} + \sum_{\{k,j\} \in
  \mathcal{E}} a_{kj} \sin(\delta_k-\delta_j) = &P_{k},  k \in \mathcal{L},
\end{align}
\end{subequations}
where equation \eqref{eq.structure-preserving1} applies at
the dynamics of generator buses and equation
\eqref{eq.structure-preserving2} applies at the dynamics of  load buses. Here $a_{kj}=V_kV_jB_{kj},$ where $B_{kj}$ is the
(normalized)  susceptance of the transmission line $\{k,j\}$ connecting the $k^{th}$ bus and $j^{th}$ bus.
$\mathcal{N}_k$ is the set of neighboring buses of the $k^{th}$
bus (see \cite{VuCCT2016} for more details). \textcolor{black}{Let $\delta(t)=[\delta_1(t)\; ...\; \delta_{|\mathcal{N}|}(t) \; \dot{\delta}_1(t) \;...\; \dot{\delta}_{|\mathcal{N}|}(t)]^\top$ be the state of the system \eqref{eq.structure-preserving} at time $t$ (for simplicity, we will denote the system state by $\delta$). Note that Eqs. \eqref{eq.structure-preserving} are invariant 
under any uniform shift of the angles $\delta_k \to \delta_k + c.$ However, the state $\delta$ can be unambiguously characterized by the
angle differences $\delta_{kj} = \delta_k-\delta_j$ and the frequencies $\dot{\delta}_k.$}

Normally, a power grid operates at an operating condition of
the \emph{pre-fault dynamics}. Under the fault, 
the system evolves according to the \emph{fault-on dynamics}. After some
time period, the fault is cleared or self-clears, and the system is at the
so-called \emph{fault-cleared state} $\delta_0$ (\textcolor{black}{the fault-cleared state is usually estimated by simulating the fault-on dynamics, and hence, is assumed to be known).} Then, the power system experiences the so-called
\emph{post-fault dynamics}.  The
transient stability assessment certifies whether the post-fault
state converges from  $\delta_0$ to a 
stable EP $\delta^*.$ Mathematically, the operating condition $\delta^*=[\delta_1^* \;...\;\delta_{|\mathcal{N}|}^* \;0 \;...\;0]^\top$ is a solution of the power-flow like equations:
\begin{align}
\label{eq.PF}
    \sum_{\{k,j\} \in
  \mathcal{E}} a_{kj} \sin\delta_{kj} = &P_{k},  k \in \mathcal{N}.
\end{align} 

\textcolor{black}{With renewable generations or under power redispatching, the power injections $P_k$ take different values. Also, the couplings $a_{kj}$ can be changed by using the FACTS devices. Assume $\underline{a}_{kj}\le a_{kj} \le \bar{a}_{kj}.$ In those situations, the resulting EP $\delta^*$ also takes different values. Therefore, we want to characterize the region of EPs so that the post-fault state always converges from a given initial state $\delta_0$ to the EP whenever the EP is in this region. Though the EP can take different values, it is assumed to be fixed in each transient stability assessment because the power injections and couplings can be assumed to be unchanged in the very fast time scale of transient dynamics (i.e., 1 to 10 seconds).} We consider the following problem: 
\begin{itemize}
\item  \textbf{Inverse \textcolor{black}{(Asymptotic)} Stability Problem:} \emph{Consider a given initial state $\delta_0.$ Assume that power injections and the line susceptances \textcolor{black}{can take different values.} Estimate the region of stable EPs \textcolor{black}{so that the state of the system} \eqref{eq.structure-preserving} always converges from $\delta_0$ to the EP in this region.}
\end{itemize}

This problem will be addressed with the inverse stability certificate to be presented in the next section.

\section{Energy function and inverse stability certificate}
\label{sec:LFF}

\subsection{Stability assessment by using energy function}
\label{stabilitycertificate}

Before introducing the inverse stability certificate addressing the inverse stability problem in the previous section, we present 
a normal stability certificate for system with the fixed power injections and line parameters.
 For the power system described by Eqs. \eqref{eq.structure-preserving}, consider the energy function: 
\begin{align}
    E\textcolor{black}{(\delta,\delta^*)}= \sum _{k \in \mathcal{G}}\frac{m_k\dot{\delta}_k^2}{2}+ \sum_{\{k,j\}\in \mathcal{E}}\int_{\delta_{kj}^*}^{\delta_{kj}}a_{kj}(\sin \xi- \sin\delta_{kj}^*)d\xi
\end{align}
Then, along \textcolor{black}{every} trajectory of \eqref{eq.structure-preserving}, we have
\begin{align}
\label{dotE}
    \dot E\textcolor{black}{(\delta,\delta^*)} &=\sum_{k\in \mathcal{G}}m_k\dot{\delta}_k\ddot{\delta}_k + \sum_{\{k,j\}\in \mathcal{E}}a_{kj}(\sin \delta_{kj}- \sin\delta_{kj}^*) \dot{\delta}_{kj} \nonumber\\
    &= \sum_{k\in \mathcal{G}}\dot{\delta}_k(P_k-d_k \dot{\delta_k} - \sum_{\{k,j\} \in
  \mathcal{E}} a_{kj} \sin(\delta_k-\delta_j)) \nonumber\\
  &+ \sum_{k\in \mathcal{L}}\dot{\delta}_k(P_k-d_k \dot{\delta_k} - \sum_{\{k,j\} \in
  \mathcal{E}} a_{kj} \sin(\delta_k-\delta_j)) \nonumber\\
  &+ \sum_{\{k,j\}\in \mathcal{E}}a_{kj}(\sin \delta_{kj}- \sin\delta_{kj}^*)(\dot{\delta}_k-\dot{\delta_j}) \nonumber\\
  &= - \sum_{k\in \mathcal{N}}d_k (\dot{\delta_k})^2 \le 0, 
\end{align}
in which the last equation is obtained from \eqref{eq.PF}. Hence,  $E(\delta,\delta^*)$ is always decreasing along \textcolor{black}{every} trajectory of \eqref{eq.structure-preserving}.

Consider the \textcolor{black}{set} $\mathcal{P}$ defined by $|\delta_{kj}| \le \pi/2, \{k,j\}\in \mathcal{E},$ 
\textcolor{black}{and the set $\Phi=\{\delta \in \mathcal{P}: E(\delta,\delta^*) <E_{\min}(\delta^*)\},$ where $E_{\min}(\delta^*)=\min_{\delta \in \partial \mathcal{P}} E(\delta,\delta^*)$ and $\partial \mathcal{P}$ is the boundary of $ \mathcal{P}.$ $\Phi$ is invariant w.r.t. \eqref{eq.structure-preserving}, and bounded as the state $\delta$ is characterized by the
angle differences and the frequencies. Though $\Phi$ is not closed, the decrease of $E(\delta,\delta^*)$ inside $\Phi$ assures the limit set to be inside $\Phi.$
As such, we can apply the LaSalle's Invariance Principle and use a proof similar to that of} Theorem 1 in \cite{VuTuritsyn:2015TAC} to show that, if $\delta_0$ is inside \textcolor{black}{$\Phi$} then the system state will only evolve inside this set and eventually converge to $\delta^*.$  
So, to check if the system state converges from $\delta_0 \in \mathcal{P}$ to $\delta^*,$ we only need to check if $E(\delta_0,\delta^*) <E_{\min}(\delta^*).$ 

\subsection{Inverse stability certificate}

\begin{figure}[h!]
\centering
\includegraphics[width = 2in]{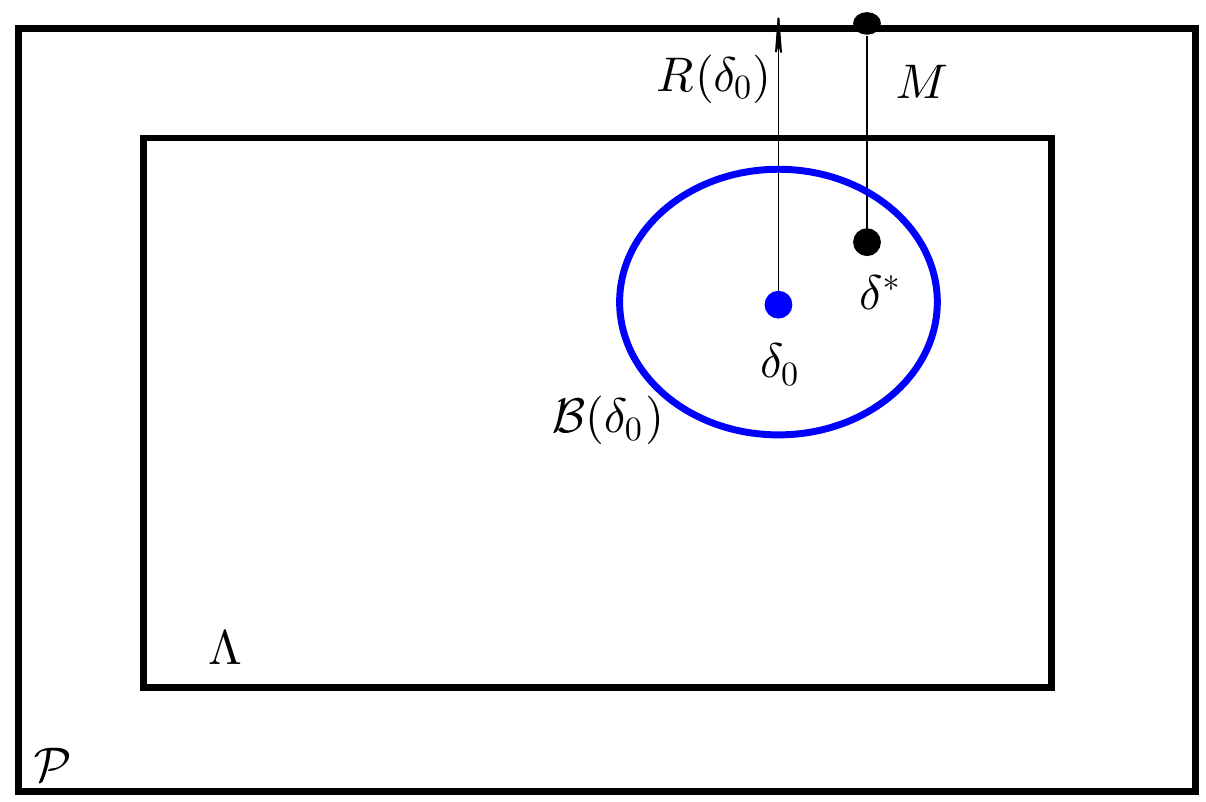}
\caption{For a power system with a given initial state $\delta_0$, if the EP $\delta^*$ is inside the set $\mathcal{A}(\delta_0)=\Lambda \cap \mathcal{B}(\delta_0)$ surrounding $\delta_0$ then the system state will converge from $\delta_0$ to the EP $\delta^*$ since $E(\delta_0,\delta^*) <E_{\min}$.} \label{fig.Inversestability}
\end{figure}

For a given initial state $\delta_0 \in \mathcal{P}$, we will construct a region surrounding it so that whenever the operating condition $\delta^*$ is in this region then $E(\delta_0,\delta^*) <E_{\min}(\delta^*).$ Hence, the grid state will converge from $\delta_0$ to $\delta^*$ according to the stability certificate in Section \ref{stabilitycertificate}.
Indeed, we establish quadratic bounds of the energy function for every $\delta^*$ in the \textcolor{black}{set} $\Lambda$ defined by inequalities $|\delta_{kj}| \le \lambda < \pi/2, \forall \{k,j\}\in \mathcal{E}.$ Let $g=\dfrac{1-\sin \lambda}{\pi/2- \lambda}>0.$ In \cite{VuTuritsyn:2015TAC}, we observed that for $\delta^*\in \Lambda,\xi\in\mathcal{P},$
\begin{align*}
 g(\xi-\delta_{kj}^*)^2 \le (\xi-\delta_{kj}^*)(\sin \xi - \sin\delta_{kj}^*) \le (\xi-\delta_{kj}^*)^2. \end{align*}
Hence, for all $\delta^*\in \Lambda,\delta\in\mathcal{P},$ we have
\begin{align}
\label{eq.lowerBound}
    E(\delta,\delta^*) &\ge \sum _{k \in \mathcal{G}}\frac{m_k\dot{\delta}_k^2}{2}+ g\sum_{\{k,j\}\in \mathcal{E}}a_{kj}\frac{( \delta_{kj}- \delta_{kj}^*)^2}{2}, \\
\label{eq.upperBound}
    E(\delta,\delta^*) &\le \sum_{k \in \mathcal{G}}\frac{m_k\dot{\delta}_k^2}{2}+ 
    \sum_{\{k,j\}\in \mathcal{E}}a_{kj}\frac{( \delta_{kj}- \delta_{kj}^*)^2}{2}.
\end{align}
 Define the following functions
\begin{align}
\label{eq.distance}
\begin{split}
    D(\delta,\delta^*) &=g\sum_{\{k,j\}\in \mathcal{E}}\underline{a}_{kj}\frac{( \delta_{kj}- \delta_{kj}^*)^2}{2}, \\
    F(\delta,\delta^*) &= \sum _{k \in \mathcal{G}}\frac{m_k\dot{\delta}_{k}^2}{2}+ \sum_{\{k,j\}\in \mathcal{E}}\bar{a}_{kj}\frac{( \delta_{kj}- \delta_{kj}^*)^2}{2}.
\end{split}
\end{align}
Using \eqref{eq.lowerBound} and \eqref{eq.upperBound}, we can bound the energy function as
\begin{align}
\label{quadraticbound}
D(\delta,\delta^*) \le E(\delta,\delta^*) \le F(\delta,\delta^*), \forall \delta\in \mathcal{P}, \delta^* \in \Lambda.
\end{align} 
For a given initial state $\delta_0$ inside \textcolor{black}{the set} $\mathcal{P},$ we calculate the ``distance'' from
this initial state to the boundary of the set $\mathcal{P}: R(\delta_0) =\min_{\delta \in \partial\mathcal{P}} D(\delta_0,\delta).$ Let $\mathcal{B}(\delta_0)$ be the neighborhood of $\delta_0$ defined by
\begin{align}
\label{setB}
  \mathcal{B}(\delta_0) =\{\delta: F(\delta_0,\delta) \le R(\delta_0)/4\}.  
\end{align}

\textcolor{black}{The following is our main result regarding inverse stability of power system}, as illustrated in Fig. \ref{fig.Inversestability}.
\emph{\begin{theorem}
\label{InverseStability}
Consider a given initial state $\delta_0$  inside the set $\mathcal{P}.$ Assume that the EP of the system \textcolor{black}{takes different values in the set $\mathcal{A}(\delta_0)=\Lambda \cap \mathcal{B}(\delta_0)$}, where the set $\mathcal{B}(\delta_0)$ is defined as in \eqref{setB}. Then, the system \textcolor{black}{state} always converges from the given initial state $\delta_0$ to the EP. 
\end{theorem}}

\emph{Proof:} See Appendix \ref{appendix}. 
$\qquad\square$


\textcolor{black}{\begin{remark}
In this paper, we limit the grid to be described by the simplified model \eqref{eq.structure-preserving} which captures the dynamics of the generators' rotor angle and the response of
load power output to frequency deviation. More realistic models should take into account voltage variations, reactive powers, 
dynamics of the rotor flux, and controllers (e.g., droop controls and power
system stabilizers). It should be noted that the results in this paper is extendable to more realistic models. The reason is that all the key results in this paper rely on the analysis of the system's energy function, in which we combine the energy function-based transient stability analysis with the quadratic bounds of the energy function. In high-order models, the energy function is more complicated, yet it still can be bounded by quadratic functions. Hence, we can combine the energy function-based transient stability analysis of the more realistic models in \cite{pai2012energy,Chiang:2011eo,Hiskens:1997Lya} with the quadratic bounds of the energy function to extend the results in this paper to these higher-order models. This approach may also extend to other higher-order models in the port-Hamiltonian formulation (e.g., \cite{port-Hamiltonian2013,Caliskan2014}) by applying the appropriate approximation on the Lyapunov functions established for these models.   
\end{remark}}

\section{Applications of Inverse Stability Certificate}

\subsection{Robust stability assessment}

The robust transient stability problem that we consider involves situations where there is
uncertainty in power injections $P_k,$ e.g., due to intermittent renewable generations. 
Formally, for a given fault-cleared state $\delta_0,$ we need to certify the transient stability of the post-fault dynamics described by \eqref{eq.structure-preserving} with respect to fluctuations of the power injections, which consequently lead to \textcolor{black}{different values} of the post-fault EP
$\delta^*$ as a solution of the power flow equations \eqref{eq.PF}. Therefore, we consider the following robust stability problem \cite{VuTuritsyn:2015TAC}:
\begin{itemize}
\item \textbf{Robust stability assessment:} \emph{Given a fault-cleared state $\delta_0,$ certify the transient stability of \eqref{eq.structure-preserving} w.r.t. a set of
stable EPs $\delta^*$ resulted from \textcolor{black}{different levels of power injections}.}
\end{itemize}

Utilizing the inverse stability certificate, we can assure robust stability of  renewable power systems whenever the resulting EP is inside the set $\mathcal{A}(\delta_0)=\Lambda \cap \mathcal{B}(\delta_0).$ 
To check that the EP is in the set $\Lambda,$ we can apply the criterion from \cite{Dorfler:2013}, which states that the EP \textcolor{black}{will be} in the set $\Lambda$ if the power injections $p=[P_1,...,P_{|\mathcal{N}|}]^T$ satisfy  
\begin{align}
\label{eq.SynchronizationCondition}
\|L^{\dag}p\|_{\mathcal{E},\infty} \le \sin\lambda,
\end{align}
where $L^\dag$ is the pseudoinverse of the network Laplacian
matrix and the norm $\|x\|_{\mathcal{E},\infty}$ is defined by
$\|x\|_{\mathcal{E},\infty}=\max_{\{i,j\}\in
\mathcal{E}}|x(i)-x(j)|.$ On the other hand, some similar sufficient condition could be developed so that we can verify that the EP is in the set $\mathcal{B}(\delta_0)$ by checking the power injections. This will help us certify robust stability of the system by only checking the power injections.

\subsection{Stability-constrained OPF}

Stability-constrained OPF problem concerns with determining the optimal operating condition with respect to the voltage and thermal constraints, as well as the stability constraint. While the voltage and thermal constraints are well modeled
via algebraic equations or inequalities, it is still
an open question as to how to include the stability constraint into OPF formulation
since stability is a dynamic concept and differential equations
are involved \cite{Pizano-Martianez2010}. 

Mathematically, a standard OPF problem is usually stated as follows \textcolor{black}{(refer to \cite{Pizano-Martianez2010} for more detailed formulation)}:
\begin{align}
    &\min c(P) \\
    \label{eqP}
    {s.t.\quad } &P(V,\delta) =P \\
    \label{eqQ}
    &Q(V,\delta) =Q \\
    \label{ineqV}
    &\underline{V} \le V \le \bar{V} \\
    \label{ineqS}
    & \underline{S} \le |S(V,\delta)| \le \bar{S}    
\end{align}
where \textcolor{black}{$c(P)$ is a quadratic cost function, the decision variables $P$ are typically the generator scheduled electrical
power outputs,} the equality constraints \eqref{eqP}-\eqref{eqQ} stand for the power flow equations, and the inequality constraints \eqref{ineqV}-\eqref{ineqS} stand for the voltage and thermal limits of
branch flows through transmission lines and transformers. Assume that the stability constraint is to make sure that the system state will converge from a given fault-cleared state $\delta_0$ to the designed operating condition, and that the reactive power is negligible. With the inverse stability certificate, the stability constraint can be relaxed and formulated as \textcolor{black}{$\delta \in \mathcal{A}(\delta_0).$} Basically, the inverse stability certificate transforms the dynamic problem of stability into a static problem of placing the prospective EP into a set. In summary, we obtain \textcolor{black}{a relaxation of the stability-constrained OPF problem as follows:}
\begin{align}
    &\min c(P) \\
    {s.t.\quad } &P(V,\delta) =P \\
    &\underline{V} \le V \le \bar{V} \\
    & \underline{S} \le |S(V,\delta)| \le \bar{S} \\
    & \delta \in \mathcal{A}(\delta_0).
\end{align}

\textcolor{black}{Solution of this optimization problem in an optimal operating condition} at which the cost function is minimized and the voltage/thermal constraints are respected. Furthermore, the stability constraint is guaranteed by the inverse stability certificate, and the system \textcolor{black}{state} is ensured to converge from the fault-cleared $\delta_0$ to the operating condition.

\subsection{Emergency control design}

\begin{figure}[h!]
\centering
\includegraphics[width = 2.4in]{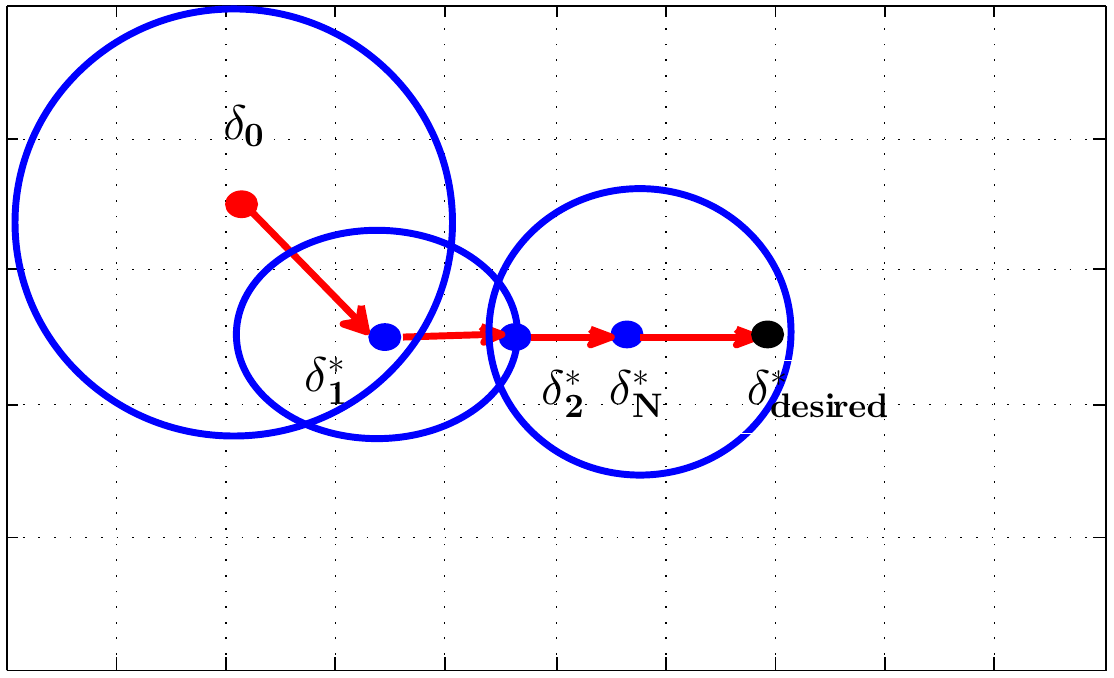}
\caption{Power dispatching to relocate the stable EPs $\delta^*_{i}$ so that the fault-cleared state, which is possibly unstable if there is no controls,
is driven through a sequence of EPs back to the desired EP $\delta^*_{\bf \textcolor{black}{desired}}$. The placement of these EPs is determined by applying the inverse stability certificate.}
\label{fig.EmergencyControl_EquilibriumSelection}
\end{figure}

Another application of the inverse stability certificate, that will be detailed in this section, is designing stability-guaranteed corrective actions that can drive the post-fault dynamics to a desired stability regime. As illustrated in Fig. \ref{fig.EmergencyControl_EquilibriumSelection}, for a given fault-cleared state $\delta_0$, by applying the inverse stability certificate, we can appropriately dispatch the power injections $P_k$ to relocate the EP of the system so that the post-fault dynamics can be attracted from the fault-cleared state $\delta_0$ through a sequence of EPs $\delta^*_1,...,\delta^*_{\ce N}$ to the desired EP $\delta^*_{\bf desired}$. \textcolor{black}{In other words, we subsequently redispatch the power injections so that the system state converges from $\delta_0$ to $\delta_1^*$, and then, from $\delta_1^*$ to $\delta_2^*,$ and finally, from $\delta_{\ce N}^*$ to $\delta^*_{\bf desired}$.} This type of corrective actions \textcolor{black}{reduces} the need for prolonged load shedding and \textcolor{black}{full} state measurement. Also, this control method is unconventional where the operating point is relocated as desired, instead of being fixed as in the classical controls.

Mathematically, we consider the following problem: 
\begin{itemize}
\item \textbf{Emergency Control Design:} \emph{Given a fault-cleared state $\delta_0$ and a desired stable EP $\delta^*_{\bf desired},$ determine the feasible dispatching of power injections $P_k$ to relocate the EPs so that the post-fault dynamics is driven from the fault-cleared state $\delta_0$ through the set of designed EPs to the desired EP $\delta^*_{\bf desired}$.}
\end{itemize}

\begin{figure}[t!]
\centering
\includegraphics[width = 2.2in]{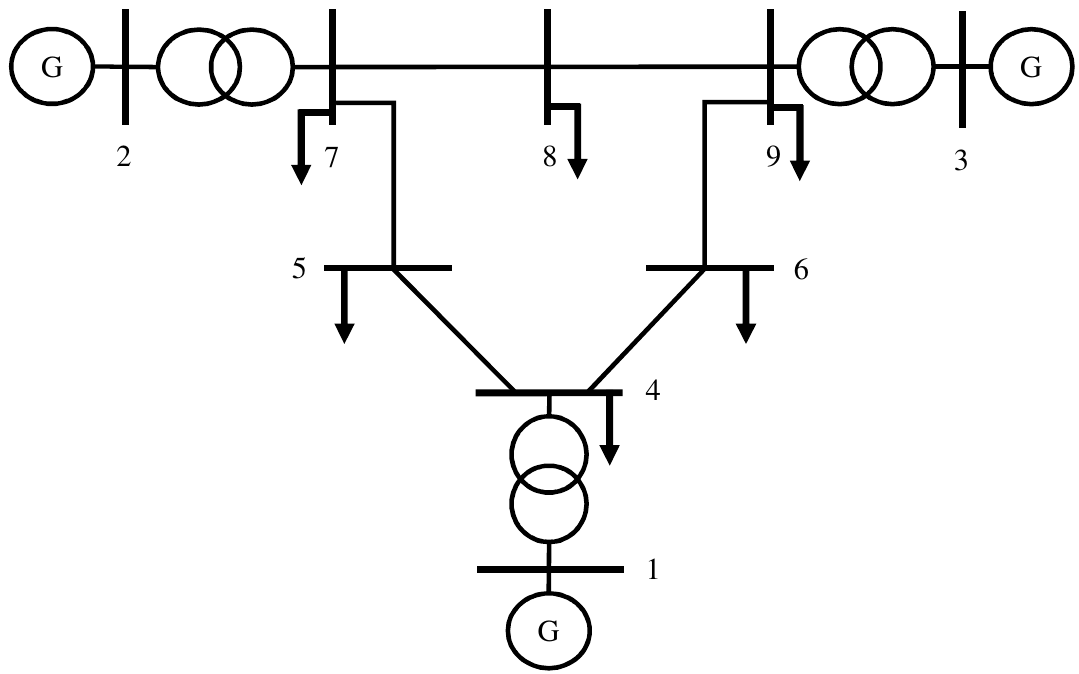}
\caption{3-generator 9-bus system with frequency-dependent dynamic
loads.} \label{fig.3generator9bus}
\end{figure}

\begin{figure}[t!]
\centering
\includegraphics[width = 2.35in]{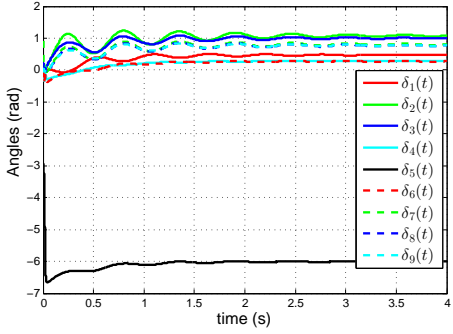}
\caption{Unstable post-fault dynamics when there is no controls: $|\delta_{45}|$ and $|\delta_{57}|$ evolve to $2\pi,$ triggering  the tripping of lines $\{4,5\}$ and $\{5,7\}$.} \label{fig.NoControl_Angle}
\end{figure}

To solve this problem, we can design the first EP $\delta^*_1$ by minimizing $\|L^{\dag}p\|_{\mathcal{E},\infty}$ over all possible power injections. The optimum power injection will
result in an EP which is 
most far away from the stability margin $|\delta_{kj}|=\pi/2,$ and hence, the stability region of  the first EP $\delta^*_1$ probably contains the given fault-cleared state.
To design the sequence of EPs, in each step, we carry out the following tasks: 

\begin{itemize}
\item Calculate the distance $R(\delta_{i-1}^*)$ from $\delta_{i-1}^*$ to the boundary of \textcolor{black}{the set} $\mathcal{P},$ i.e., $R(\delta_{i-1}^*) =\min_{\delta \in \partial\mathcal{P}} D(\delta,\delta_{i-1}^*).$ Noting that minimization of $D(\delta,\delta_{i-1}^*)$ over the boundary of the set $\mathcal{P}$ is a convex problem with a quadratic objective function and linear constraints. Hence, we can quickly obtain  $R(\delta_{i-1}^*)$.
\item Determine the set $\mathcal{B}(\delta_{i-1}^*)$ and the set $\mathcal{A}(\delta_{i-1}^*).$
\item The next EP $\delta_i^*$ will be chosen as the intersection of the boundary of the set $\mathcal{A}(\delta_{i-1}^*)$ and the line segment connecting $\delta_{i-1}^*$ and $\delta_{\bf desired}^*.$
\item The power injections $P_k$ that we have to redispatch will be determined by $P_k^{(i)}=\sum_{j\in \mathcal{N}_k} a_{kj}\sin \delta_{i_{kj}}^*$ for all $k.$ This power dispatch will place the new EP at $\delta^*_i$ which is in the inverse stability region of the previous EP $\delta_{i-1}^*$. Therefore, the controlled post-fault dynamics will converge from $\delta_{i-1}^*$ to $\delta_i^*$. 
\end{itemize}  

This procedure strictly reduces the distance from EP to $\delta_{\bf desired}^*$ \textcolor{black}{(it can be proved that there exists a constant $d>0$ so that such distance reduces at least $d$ in each step)}. Hence, after some steps, the EP $\delta_{\ce N}^*$ will be sufficiently near the desired EP $\delta_{\bf desired}^*$ so that the convergence of the system state to the desired EP $\delta_{\bf desired}^*$ will be guaranteed. 

\begin{table}[ht!]
\centering
\begin{tabular}{|c|c|c|}
  \hline
  Node & V (p.u.) & $P_k$ (p.u.) \\
  \hline
  1 & 1.0284 & 3.6466 \\
  2 & 1.0085 & 4.5735 \\
  3 & 0.9522 &  3.8173 \\
  4 & 1.0627 & -3.4771 \\
  5 & 1.0707 & -3.5798 \\
  6 & 1.0749 & -3.3112 \\
  7 & 1.0490 & -0.5639 \\
  8 & 1.0579 &  -0.5000 \\
  9 & 1.0521 &  -0.6054 \\
  \hline
\end{tabular}
\caption{Bus voltages, mechanical inputs, and static
loads.}\label{tab.data9bus}
\end{table}

\begin{figure}[t!]
\centering
\includegraphics[width = 2.4in]{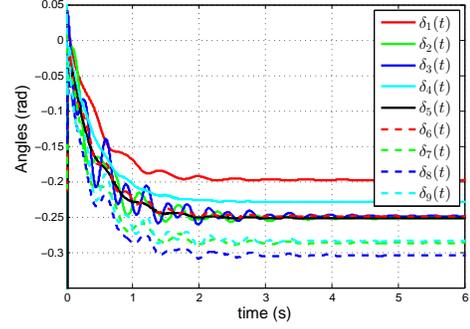}
\caption{Stable dynamics with power injection control: Convergence of buses angles from the fault-cleared state to $\delta_1^*$ in the post-fault dynamics} \label{fig.InjectionControl_Angle}
\end{figure}
\begin{figure}[t!]
\centering
\includegraphics[width = 2in]{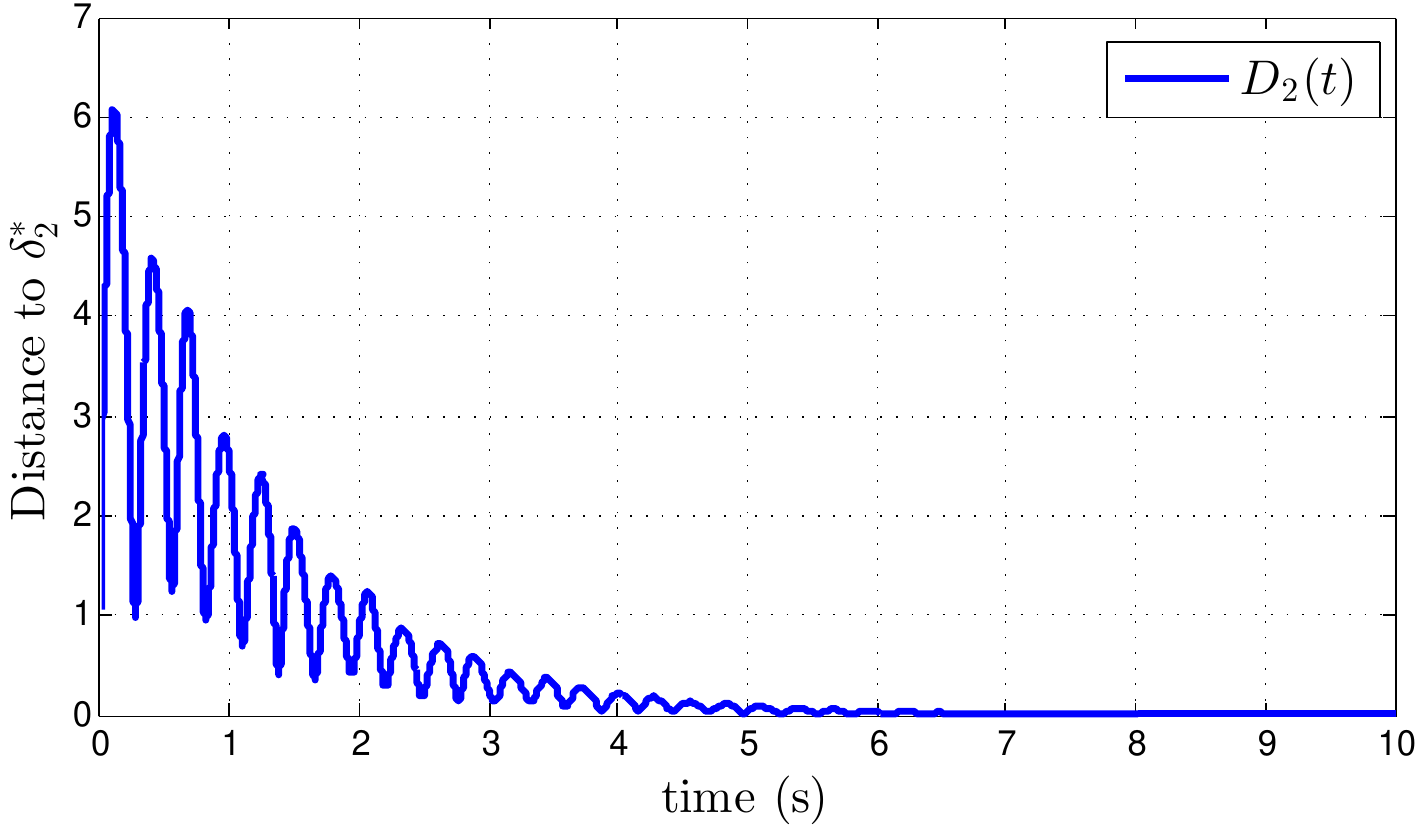}
\caption{Effect of power dispatching control: the convergence of the distance $D_2(t)$ to $0$. Here, the Euclid distance $D_2(t)$ between a state $\delta$ and the second EP $\delta_2^*$
is defined as $D_2(t)=\sqrt{\sum_{i=2}^{9} (\delta_{i1}(t)-\delta_{2_{i1}}^*)^2}$.} \label{fig.PowerDispatching_Distance}
\end{figure}

\begin{figure}[t!]
\centering
\includegraphics[width = 2in]{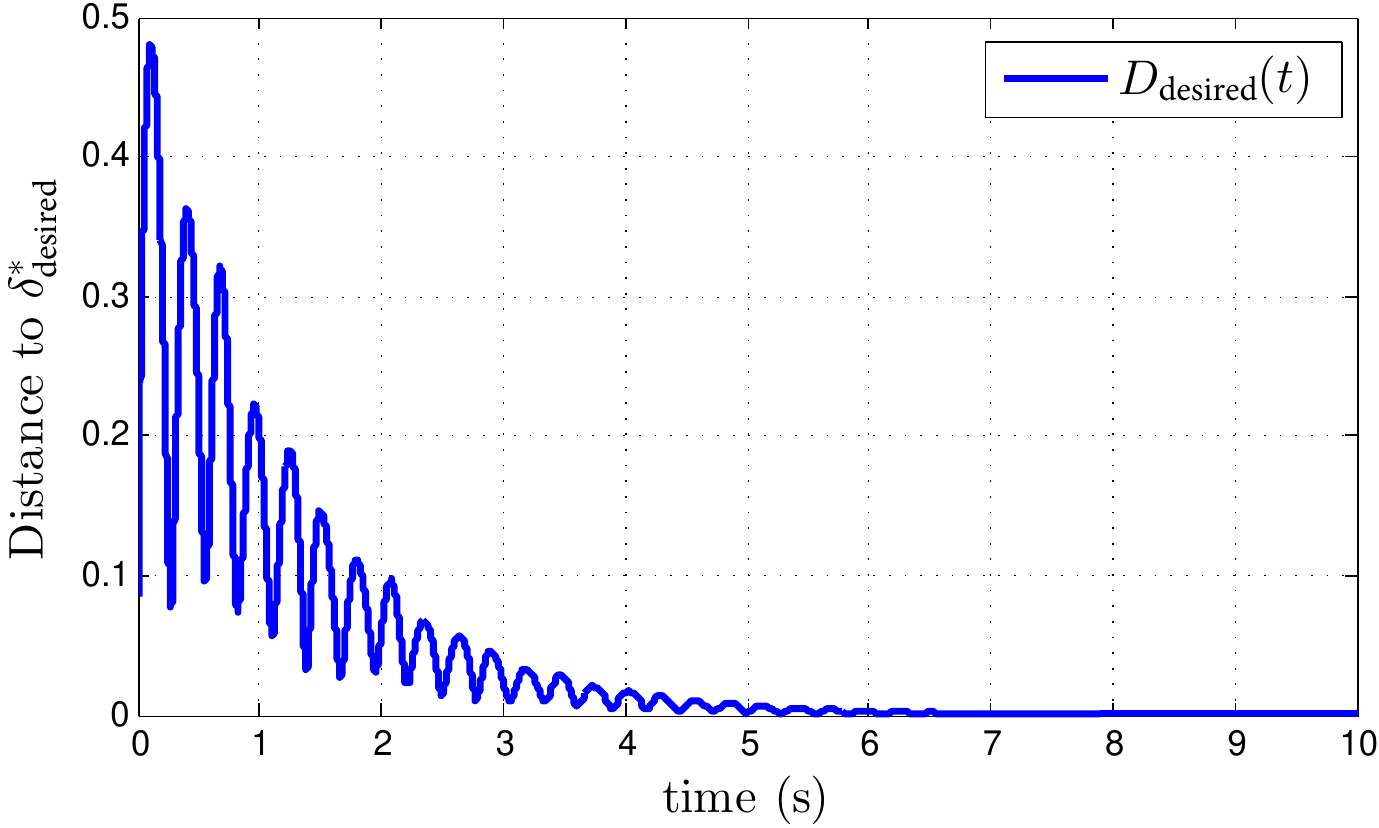}
\caption{Autonomous dynamics when we switch the power injections to the desired values: the convergence of the distance $D_{\bf \textcolor{black}{desired}}(t)$ to $0$. Here, the distance $D_{\bf \textcolor{black}{desired}}(t)$ between a state $\delta$ and the desired EP $\delta_{\bf desired}^*$
is defined as $D_{\bf desired}(t)=\sqrt{\sum_{i=2}^{9} (\delta_{i1}(t)-\delta_{{\bf desired}_{i1}}^*)^2}$.} \label{fig.Autonomous_Distance}
\end{figure}

To illustrate that this control works well in stabilizing some possibly unstable fault-cleared state $\delta_0$, we consider the 3-machine 9-bus system with 3 generator buses and 6 frequency-dependent load buses as in Fig. \ref{fig.3generator9bus}.
The susceptances of the transmission lines are as follows:
$B_{14}=17.3611 p.u.,B_{27}=16.0000 p.u.,B_{39}= 17.0648 p.u.,
B_{45}=11.7647 p.u., B_{57}= 6.2112p.u., B_{64}=10.8696p.u.,
B_{78}= 13.8889p.u.,B_{89}=9.9206p.u., B_{96}=5.8824p.u.$ 
    The parameters for generators are: $m_1=0.1254, m_2=0.034, m_3=0.016, d_1=0.0627, d_2=0.017, d_3=0.008.$ For simplicity, we take $d_k=0.05, k=4\dots,9.$  Assume that the fault trips the line between buses $5$ and $7,$ and make the power injections to fluctuate. When the fault is cleared this line is re-closed. We also assume the fluctuation of the generation (probably due to renewables) and load so that the voltages $V_k$ and power injections $P_k$ of the post-fault dynamics are given in Tab. \ref{tab.data9bus}. The stable
EP $\delta_{\bf desired}^*$ is calculated 
 as $[-0.1629\;
    0.4416\;
    0.3623\;
   -0.3563\;
   -0.3608\;
   -0.3651\;
    0.1680\;
    0.1362\;
    0.1371]^\top.$ However,  the fault-cleared state, with angles
   $[0.025 \;-0.023\; 0.041\; 0.012\; -2.917\; -0.004\; 0.907\; 0.021\; 0.023]^\top$ and generators angular velocity $[-0.016\; -0.021\; 0.014]^\top,$  is outside the set $\mathcal{P}.$
It can be seen from Fig. \ref{fig.NoControl_Angle} that the uncontrolled post-fault dynamics is not stable since $|\delta_{45}|$ and $|\delta_{57}|$ quickly evolve from initial values to $2\pi,$ which will activate the protective devices to trip the lines.

Using CVX software \textcolor{black}{\cite{grantcvx}} to minimize $\|L^{\dag}p\|_{\mathcal{E},\infty},$ we obtain the new power injections at buses 1-6 as follows: $P_1= 0.5890, P_2=
    0.5930, P_3=
    0.5989, P_4=
   -0.0333, P_5=
   -0.0617,$ and $P_6=
   -0.0165.$ Accordingly,
the minimum value of $\|L^{\dag}p\|_{\mathcal{E},\infty}=0.0350 < \emph{\emph{sin}}(\pi/89).$ Hence, the first EP obtained from equation \eqref{eq.PF} will be in the set defined by the inequalities $|\delta_{kj}|\le \pi/89, \forall \{k,j\}\in\mathcal{E},$ and can be approximated by
$\delta^*_1 \approx L^{\dag}p=[  0.0581\;
    0.0042\;
    0.0070\;
    0.0271\;
    0.0042\;
    0.0070\;
   -0.0308\;
   -0.0486\;
   -0.0281]^\top$. The simulation results confirm that the post-fault dynamics is made stable by applying the optimum power injection control, as showed in Fig. \ref{fig.InjectionControl_Angle}. Using the above procedure, after one step, we can find that $\delta_2^*=0.9259 \delta^*_{\bf desired}+0.0741\delta_1^*$ is the intersection of the set \textcolor{black}{$\mathcal{A}(\delta_1^*)$} and the line segment connecting $\delta_1^*$ and $\delta^*_{\bf desired}.$ This EP is inside the inverse stability region of $\delta_1^*,$ and hence the system state will converge from $\delta_1^*$ to $\delta_2^*$ when we do the power dispatching $P_k^{(2)}$ corresponding to $\delta_2^*$. On the other hand, $\delta_2^*$ is very near the desired EP $\delta^*_{\bf desired}$ and it is easy to check that $\delta^*_{\bf desired}$ is in the inverse stability region of $\delta_2^*,$ and thus the system state will converge from $\delta_2^*$ to the desired EP  $\delta^*_{\bf desired}$. Such convergence of the controlled post-fault dynamics is confirmed in Figs. \ref{fig.PowerDispatching_Distance}-\ref{fig.Autonomous_Distance}.

\section{Conclusions}
\label{sec.conclusion}

Electric power grids possess rich dynamical behaviours, e.g., nonlinear interaction, prohibition of global stability, and exhibition of significant uncertainties, that challenge the maintenance of their reliable operation and pose interesting questions to control and power communities. This letter characterized a surprising property termed as ``inverse stability'', which was rarely investigated
and poorly understood \textcolor{black}{(though some related inverse problems were addressed in \cite{Hiskens2004})}. 
This new notion could change the way we think about the stability assessment problem. Instead of estimating the set of initial states leading to a given operating condition, we characterized the set of operating conditions that a power grid converges to from a given initial state under changes in power injections and lines.
In addition, we briefly described three applications of the inverse stability certificate: (i) assessing the stability of renewable power systems, (ii) solving the stability-constrained OPF problem, and (iii) designing power dispatching remedial actions to recover the transient stability of power systems. Remarkably, we showed that robust stability due to the fluctuation of renewable generations can be effectively assessed, and that the stability constraint can be \textcolor{black}{incorporated  as a static constraint into the conventional OPF.}
We also illustrated a unconventional control method, in which we appropriately relocate the operating point to attract a given fault-cleared state that originally leads to an unstable dynamics. 

\section{Appendix: Proof of Theorem 1}
\label{appendix}

For each EP $\delta^* \in \mathcal{A}(\delta_0)$, let $M$ be the point on the boundary of \textcolor{black}{the set $\mathcal{P}$ so that} $E(M,\delta^*) =  E_{\min}(\delta^*)=\min_{\delta \in \partial\mathcal{P}}E(\delta,\delta^*),$ as showed in Fig. \ref{fig.Inversestability}. From \eqref{quadraticbound}, we have
\begin{align}
\label{eq.20}
    &E(M,\delta^*) + E(\delta_0,\delta^*) \ge D(M,\delta^*) + D(\delta_0,\delta^*) \nonumber \\
     &= g\sum_{\{k,j\}\in \mathcal{E}}\underline{a}_{kj}\frac{( \delta_{M_{kj}}- \delta_{kj}^*)^2+ ( \delta_{0_{kj}}- \delta_{kj}^*)^2}{2} \nonumber \\
     &\ge g\sum_{\{k,j\}\in \mathcal{E}}\underline{a}_{kj}\frac{(\delta_{M_{kj}}- \delta_{0_{kj}})^2}{4} =\frac{D(\delta_0,M)}{2}.
\end{align}
Note that $D(\delta_0,M) \ge R(\delta_0)$ as $R(\delta_0)$ is the distance from $\delta_0$ to the boundary of \textcolor{black}{the set} $\mathcal{P}.$   This, together with \eqref{eq.20}, leads to $E(M,\delta^*) + E(\delta_0,\delta^*) \ge R(\delta_0)/2.$ From \eqref{quadraticbound} and \eqref{setB}, we have 
$E(\delta_0,\delta^*) \le F(\delta_0,\delta^*) < R(\delta_0)/4.$
Hence, 
\begin{align}E(M,\delta^*) > R(\delta_0)/4 > E(\delta_0,\delta^*).\end{align} 
Therefore, for any $\delta^* \in \mathcal{A}(\delta_0),$ we have
$E(\delta_0,\delta^*) < E_{\min}(\delta^*).$ \textcolor{black}{By applying the stability analysis in Section \ref{stabilitycertificate}, we conclude that} the initial state $\delta_0$ must be inside the stability region of the EP $\delta^*$ and the system \textcolor{black}{state} will converge from the initial state $\delta_0$ to the EP $\delta^*.$

\textcolor{black}{\section{Acknowledgements}
This work was supported by the MIT/Skoltech, Ministry of Education and Science of Russian Federation (Grant no. 14.615.21.0001.), and NSF under Contracts 1508666 and 1550015.}

\bibliographystyle{IEEEtran}
\bibliography{lff}
\end{document}